\documentclass[runningheads]{llncs}
\usepackage[T1]{fontenc}
\usepackage{graphicx,verbatim}
\usepackage{amsmath}
\usepackage{amssymb}
\usepackage{subcaption}
\usepackage{multirow}

\usepackage{color}
\begin{document}
\title{FedProIn: Mitigating Client Drift for Learnable Prototypes in Federated Medical Imaging}
\begin{comment}  %% Removed for anonymized MICCAI submission
\author{First Author\inst{1}\orcidID{0000-1111-2222-3333} \and
Second Author\inst{2,3}\orcidID{1111-2222-3333-4444} \and
Third Author\inst{3}\orcidID{2222--3333-4444-5555}}
%
\authorrunning{F. Author et al.}
\institute{Princeton University, Princeton NJ 08544, USA \and
Springer Heidelberg, Tiergartenstr. 17, 69121 Heidelberg, Germany
\email{lncs@springer.com}\\
\url{http://www.springer.com/gp/computer-science/lncs} \and
ABC Institute, Rupert-Karls-University Heidelberg, Heidelberg, Germany\\
\email{\{abc,lncs\}@uni-heidelberg.de}}

\end{comment}

\author{Harsh Kumar\inst{1,3} \and Tarun Kumar Garg \inst{2,3} \and Vaanathi Sundaresan\inst{1,4}}

% index{Kumar, Harsh}
% index{Garg, Tarun Kumar}
% index{Sundaresan, Vaanathi}

\institute{Department of Computational and Data Sciences,
Indian Institute of Science\\
Bengaluru, Karnataka 560012, India\\ \and
Department of IISc Mathematics Initiative,
Indian Institute of Science\\
Bengaluru, Karnataka 560012, India\\ \and
Equally contributed to the work \and
Corresponding author:
\email{vaanathi@iisc.ac.in}
}
  
\maketitle          % typeset the header of the contribution
\begin{abstract}

Federated learning (FL) is severely hindered by statistical heterogeneity due to variations in scanners, acquisition protocols, and patient populations. Such non-IID data induces client drift during local optimization, leading to unstable convergence and suboptimal global models when parameter-based aggregation is applied. We propose a prototype-based, influence-aware federated learning framework (FedProIn) that uses multiple learnable class prototypes to capture shared semantic structures across heterogeneous clients. We introduce feature divergence loss and prototype contrastive loss to mitigate client drift by decomposing it into feature drift and prototype drift.
In addition, we propose a normalized influence aggregation strategy that adaptively weights client prototypes according to their contribution to the global representation, reducing the impact of biased or low-quality updates. Experimental results on two publicly available medical datasets, HAM10000 and Matek-19, demonstrate that FedProIn achieves accuracies of (83.5\% IID, 81.1\% non-IID) on HAM10000 and (96.2\% IID, 95.8\% non-IID) on Matek-19, respectively,  outperforming existing baselines in both conditions. Our code is available at \url{https://github.com/harsh-kmr/FedProIn}.

%The abstract should briefly summarize the contents of the paper in 150--250 words.  If you are to include a link to your Repository, please make sure it is anonymized for the double-blind review phase. 

% Authors must provide keywords and are not allowed to remove this Keyword section.

\end{abstract}
\section{Introduction}
Medical imaging research faces limitations posed by small datasets and shortage of resources~\cite{guan2024fl}. Further,  pooling data across centers is limited by privacy regulations like HIPAA~\cite{hipaa} and GDPR~\cite{gdpr}. Federated Learning (FL) allows local models to be trained on-site while sharing only updates with a central server. However, the performance of FL is often compromised by statistical heterogeneity in medical deployments, due to differences in scanner protocols, demographics, and disease prevalence \cite{guan2024fl}. This heterogeneity causes local models to diverge from the global model, leading to \textit{client drift} and degraded FL performance. This is mainly due to heterogeneous data distributions leading to identical instances getting mapped by different client models to distinct regions in the feature space. Additionally, local representations also often become biased toward client-specific data characteristics, causing them to deviate from the global representation.

\paragraph{\textbf{Related work.}} Recent literature highlights two approaches. The first, constraint-based regularization, includes methods like FedProx \cite{fedprox}, and MOON \cite{moon}, which add penalties to limit disparity between local and global model parameters. Though effective, these methods struggle to align semantic representations in high-dimensional space. The second is prototype-based learning ~\cite{fedproto,fedproc,fedtgp,fpl,fedplvm,fedpcl}, using prototypes to guide local training. Typically, methods aggregate these prototypes as mean feature vectors, summarizing each class by a single global prototype via weighted averaging of local prototypes. Most approaches (e.g., FedProto \cite{fedproto}, FedProc \cite{fedproc}) assume each class is represented by a single prototype, which is inadequate in medical FL due to data heterogeneity. Some methods use multiple prototypes (e.g., FedPLVM \cite{fedplvm}, FPL \cite{fpl}), but clustering adds complexity. Methods like FedPCL \cite{fedpcl} use multiple encoders for distinct prototypes, increasing costs. FedTGP \cite{fedtgp} introduces trainable global prototypes optimized on servers, but still relies on unimodal client representations, using mean feature vectors. These prototypes often sacrifice inter-class separability by not optimizing decision boundaries. Learnable prototypes (LPs) maximize class separation by maximizing the margin between prototypes, indicating stronger separability. 
Convolutional Prototype Learning (CPL)~\cite{yang_cvpr_2018} showed that optimizing learnable prototypes and a feature extractor from raw data creates intra-class compact and inter-class separable representations, versus using a softmax layer. However, this remains unexplored in FL setup.

\begin{comment}
\begin{figure}
    \centering
    % First Subfigure (a)
    \begin{subfigure}[b]{0.32\textwidth}
        \centering
        \includegraphics[width=\linewidth]{margin_prototypes.pdf}
        \caption{Inter-Class Separability}
        \label{fig:separability}
    \end{subfigure}
    \hfill % Adds flexible space between images
    % Second Subfigure (b)
    \begin{subfigure}[b]{0.32\textwidth}
        \centering
        \includegraphics[width=\linewidth]{feature_drift_expExperiment5_Loner_DIRICHLET_NUMPROTOS1_c0_vs_c1_class0.pdf}
        \caption{Feature Drift}
        \label{fig:Feature_drift}
    \end{subfigure}
    \hfill
    % Third Subfigure (c)
    \begin{subfigure}[b]{0.32\textwidth}
        \centering
        \includegraphics[width=\linewidth]{proto_drift_expExperiment5_Loner_DIRICHLET_NUMPROTOS1_class0.pdf}
        \caption{Prototype Drift}
        \label{fig:Prototype_drift}
    \end{subfigure}
    
    \caption{(a) The heatmaps show pairwise Euclidean distances between class prototypes for FedProIn (lower) and FedProto (upper), with the margin defined as the minimum distance between distinct prototypes \((\min_{i \ne j}\|p_i - p_j\|_2)\). (b) UMAP visualization of feature vectors for 500 identical images processed by two distinct client models of training learnable prototype without federation on HAM10000 dataset. (c) UMAP projection of Class 0 learnable prototypes across 10 clients from same experiment.}
    \label{fig:three_images}
\end{figure}
\end{comment}
\paragraph{\textbf{Contributions.}} In this work, we propose \textbf{Fed}erated Learning with \textbf{Pro}totype aggregation with \textbf{In}fluence (\textbf{FedProIn}) to maximize class separation. Firstly, we adopt the LP formulation that treats prototypes as optimizable model parameters, unlike standard approaches that rely on mean vectors. Secondly, we explicitly tackle client drift by decomposing it into feature drift and prototype drift, addressing them separately using feature divergence loss and prototype contrastive loss. Additionally, we also introduce a Normalized Influence Aggregation (NIA) mechanism that efficiently aggregates multi-prototype models by quantifying the adaptive relevance of prototypes in each client. We evaluate our approach against existing methods on public long-tailed medical imaging datasets. 

\section{Methodology}
Our objective is to train a global model that maps an input image $ x \in \mathcal{X} $ to a label $ y \in \{1,\dots,C\} $, where $ C $ denotes the number of classes. Unlike standard classification approaches that map inputs directly to logits, we adopt a prototype-based learning framework where the global model is decomposed into two distinct components: a feature encoder $ f_\theta $ parameterized by $ \theta $, which maps input images to a $ d $-dimensional embedding space, and a set of LPs $ \mathcal{P} \in \mathbb{R}^{C \times M \times d} $, where $ M $ denotes the number of prototypes per class. The global model parameters are denoted as $ w = \{\theta, \mathcal{P}\} $.

We consider a setup with $ K $ clients, where each client \( i \in \{1,\dots,K\} \) owns a private local dataset \( \mathcal{D}_i = \{(x_{i,j}, y_{i,j})\}_{j=1}^{n_i} \). Here, $ n_i = |\mathcal{D}_i| $ is the size of the local dataset, and $ n = \sum_{i=1}^{K} n_i $ denotes the total number of samples across all clients. The goal is to minimize the global empirical risk, which is formulated as
\begin{equation}
\min_{w} \; \mathcal{L}(w) = \sum_{i=1}^{K} \frac{n_i}{n} \ell(w, \mathcal{D}_i ), 
\end{equation}
where \( \ell(w, \mathcal{D}_i) \) is 
 the local objective at the client $i$ over the client's dataset $\mathcal{D}_i$.

\subsection{Federated Learning with Prototype aggregation with Influence (FedProIn)} 
\begin{figure}
    \centering
    \includegraphics[width=1\linewidth]{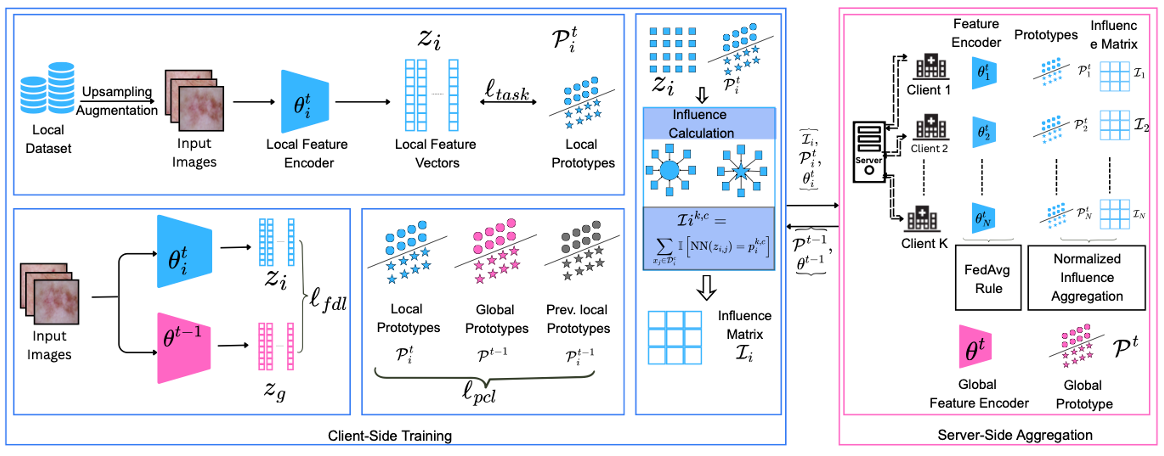}
    \caption{\textbf{Overview of FedProIn framework.} In client-side training (blue), clients extract features using local encoder $\theta_i^t$ and update local prototypes $\mathcal{P}_i^t$. \emph{Feature Divergence Loss (FDL)} aligns local and frozen global features, while \emph{Prototype Contrastive Loss (PCL)} pulls local prototypes toward global prototypes and pushes away from previous ones (grey). Clients construct an influence matrix $\mathcal{I}_i$ tracking prototype usage. In server-side aggregation (pink), encoders are aggregated via FedAvg and prototypes via \emph{Normalized Influence Aggregation (NIA)} to produce global encoder $\theta^t$ and prototypes $\mathcal{P}^t$.}
    \label{fig:fedproin}
\end{figure}

In FedProIn, we consider prototypes as model parameters~\cite{yang_cvpr_2018}, instead of mean feature vectors, optimized jointly with the feature extractor via backpropagation. Fig.~\ref{fig:fedproin} illustrates the overall framework of FedProIn. At the start of the communication round $ t $, the server broadcasts the global state $ w^{t-1} $ to the participating clients. Clients perform local training and return the updated parameters $ w^{t}_i $ alongside an \textit{influence matrix} $ \mathcal{I}_i $, which guides the server's aggregation.

\paragraph{\textbf{\textit{Client-side local training.}}}
The local training process addresses client drift through: (i) \textit{feature drift}, where client models map images to different regions in feature space due to data heterogeneity, and (ii) \textit{prototype drift}, where client prototypes are pulled towards local data characteristics, causing prototypes to diverge from global representation. These lead to semantic misalignment, degrading global model robustness. Hence, we introduce two regularization terms.

\paragraph{\underline{Feature divergence loss (FDL).} } To address feature drift, we employ a mean square loss between the global and local feature vectors. Each client maintains a copy of the global feature extractor (as shown in Fig.~\ref{fig:fedproin}), parameterized by $ \theta^{t-1} $, received from the server at the start of round $t$. This frozen encoder is used to compute reference global feature representations and is not involved in local training. Given \( z_{i,j} = f_{\theta_i^t}(x_{i,j}) \) and \( z_{g,j} = f_{\theta^{(t-1)}}(x_{i,j}) \), we align the local features toward the global features by minimizing:
\begin{equation}
\ell_{FDL} = \frac{1}{n_i} \sum_{j=1}^{n_i} \| z_{i,j} - z_{g,j} \|_2^2
\end{equation}
 \underline{\textit{Prototype contrastive loss (PCL).} } To regulate prototype drift, we utilize a contrastive loss since global prototypes likely capture more generalized representation of the overall data distribution compared to local prototypes that might be skewed due to shift in their data distribution. Let $\mathcal{P}^{(t-1)}$ denote the current global prototypes and $\mathcal{P}_i^{(t-1)}$ denote the client's prototypes from the previous round. We aim to simultaneously pull the updated local prototypes $\mathcal{P}_{i}^{t}$ closer to the global prototypes while pushing them away from their previous states $\mathcal{P}_i^{(t-1)}$. To achieve this, we compute distances $d_{\text{glob}} = \|\mathcal{P}^{(t-1)} - \mathcal{P}_i^t\|_2$ and $d_{\text{prev}} = \|\mathcal{P}_i^{(t-1)} - \mathcal{P}_i^t\|_2$, and minimize:
\begin{equation}
\ell_{PCL} = \max(0, \Delta + m), \quad \text{where} \quad \Delta = \frac{d_{\text{glob}} - d_{\text{prev}}}{d_{\text{glob}} + d_{\text{prev}} + \epsilon} 
\end{equation}
where $m$ is a margin hyperparameter (so that $\mathcal{P}_{i}^{t}$ must be closer to $\mathcal{P}^{(t-1)}$ than to $\mathcal{P}_i^{(t-1)}$ by at least $m$) and $\epsilon$ is a stability constant in the denominator. 

\paragraph{\underline{The Update Challenge and Class Imbalance.} }Optimizing multiple prototypes creates a structural \textit{update challenge} in contrastive loss. For any sample, the loss function interacts with the nearest true-class prototype (positive) and competing-class prototype (negative), with gradients flowing to these two prototypes while others receive zero gradients. In long-tailed medical datasets, prototypes for rare disease subtypes may rarely be selected as nearest neighbors, causing them to stagnate and fail to represent their semantic subregions. To address this, we use class-balanced sampling and adaptive augmentation~\cite{balancefl}, where clients oversample minority classes and apply augmentation to generate diverse views for rare subtypes. 

The total local objective for client $i$ combines the primary classification loss $\ell_{Task}$ with FDL and PCL terms as follows: 
\begin{equation}
\ell_{\text{Total}} = \ell_{Task} + \lambda_{FDL} \ell_{FDL} + \lambda_{PCL} \ell_{PCL},
\end{equation}
where $\lambda_{FDL}$ and $\lambda_{PCL}$ are weighting coefficients of $\ell_{FDL}$ and $\ell_{PCL}$, respectively.  

\paragraph{\textit{\textbf{Server-side Aggregation.}}}
At the end of round $t$, the server receives updated model parameters $ \{\theta_i^{t}, \mathcal{P}_i^{t}\} $ and influence matrix $ \mathcal{I}_i $ from each client $ i \in \mathcal{S}_t $, where $ \mathcal{S}_t $ represents participating clients. The server coordinates global update by aggregating encoder parameters and prototypes via distinct protocols, as described below. 

\paragraph{\underline{Encoder parameter aggregation.} }
We employ the standard FedAvg protocol~\cite{fedavg} to aggregate the feature encoder. The global encoder is computed as a weighted average based on local dataset sizes and $N_t = \sum_{i \in \mathcal{S}_t} n_i$.
\begin{equation}
\theta^{t} = \sum_{i \in \mathcal{S}_t} \frac{n_i}{N_t} \theta_i^{t},
\end{equation}
\underline{\textit{Normalized influence aggregation (NIA) of prototypes.} } Unlike the encoder, FedAvg/ FedProto aggregation cannot be directly applied to prototypes due to limitations with multiple prototypes per class: (1) \textit{Heterogeneous prototype utilization:} within a class, prototypes are not utilized uniformly; some are more frequently activated than others. Hence, FedAvg assigns prototype weights proportional to the dataset size $ n_{i} $, amplifying noise from less utilized prototypes. (2) \textit{Sparse Prototype Activation:} Some prototypes remain unutilized by specific clients when representing rare disease subtypes are absent from local distribution.  FedAvg would assign them the full $ n_{i} $ weight, suppressing information from clients that utilize these prototypes, potentially losing rare but important case representations.

To address these limitations, we propose NIA mechanism. During the local training phase at client $ i $, we explicitly track the utility of each prototype by constructing an influence matrix $ \mathcal{I}_i $. For the $ k $-th prototype of class $ c $, denoted as $ p_{i}^{k,c} $, the corresponding influence score $ \mathcal{I}_{i}^{k,c} $ is computed as:
\begin{equation}
\mathcal{I}_i^{k,c} = \sum_{x_j \in \mathcal{D}_i^c} \mathbb{I}\left[\text{NN}(z_{i,j}) = p_i^{k,c}\right],
\end{equation}
where $\mathcal{D}_i^c$ is the set of training samples from class $c$ at client $i$, $z_{i,j}$ is the feature embedding of sample $x_j$,  $\text{NN}(\cdot)$ returns the nearest prototype (in Euclidean distance), and $\mathbb{I}[\cdot]$ is the indicator function.
Intuitively, $\mathcal{I}_i^{k,c}$ counts how many times prototype $p_i^{k,c}$ served as the closest match for local samples during training.

The server then aggregates the prototypes as follows. For each class $ c $ and prototype index $ k $, the global prototype is computed as:
\begin{equation}
p_{global}^{k,c,t} = \frac{\sum_{i \in \mathcal{S}_t} \mathcal{I}_i^{k,c} \cdot p_i^{k,c}}{\sum_{i \in \mathcal{S}_t} \mathcal{I}_i^{k,c} + \epsilon}
\end{equation}
This formulation weights prototypes proportionally to their usage during local training, automatically down-weighting noisy updates from underutilized prototypes (addressing limitation 1) while preserving specialized prototypes relevant to a subset of clients (addressing limitation 2).

\section{Experimental setup}
\textit{\textbf{Dataset details.}} We evaluate FedProIn on two public long-tailed medical imaging datasets: (i) HAM\-10000 \cite{ham10000} contains 10,015 dermatoscopic images of pigmented skin lesions collected from multiple sources with 7 classes, and (2) Matek-19 \cite{Matek-19} consists of over 18,000 annotated peripheral blood cell images collected from 100 patients diagnosed with acute myeloid leukemia. 
All images are resized to $128 \times 128$ pixels and normalized using ImageNet statistics \cite{imagenet} prior to training. For both datasets, the training:validation:test split is 70\%:10\%:20\% respectively. \textit{Federated Simulation:} We simulate FL setup with $K=10$ clients under two partitioning schemes: (1) \textit{IID}, and (2) \textit{Non-IID}, using a Dirichlet distribution with $\alpha=0.5$ to model statistical heterogeneity.

\paragraph{\textbf{Implementation.}} All experiments use a pretrained ResNet-18 backbone, trained using the Adam optimizer \cite{adam} (with a learning rate of $1\times10^{-4}$ and a batch size of 128) for 50 communication rounds with 50\% client participation per round. The parameters are set empirically. Results are averaged over three independent runs. For federated methods, the number of local epochs is set to $E=5$. For $\ell_{task}$, we used Minimum Classification Error Loss (MCEL) and Margin-based Classification Loss (MCL) {\cite{yang_cvpr_2018}}. For MCEL, for FDL and PCL, the corresponding values of $\lambda$ are set to 1.0, and for MCL, $\lambda_{FDL}$ and $\lambda_{PCL}$ are 0.1 and 0.01, respectively for both datasets. The values are determined empirically. The implementation is done using PyTorch \cite{paszke2019pytorch} (2.6.0). Experiments are conducted on an Intel i9-14900k CPU, 64 GB RAM, and one NVIDIA RTX A5000 GPU ( with 24 GB memory). 

\paragraph{\textbf{Evaluation Metrics. }}
We evaluate classification performance using: Accuracy (Acc) in \%, Weighted F1-Score (W-F1) in \%, and Matthews Correlation Coefficient (MCC) \cite{mcc}. 
MCC computes the Pearson correlation coefficient between the observed and predicted classifications. MCC yields a value between -1 and +1, offering a balanced metric even when classes are of very different sizes.

\subsection{Experiments and results}
\textit{\textbf{Comparison with state-of-the-art methods. }}We initially determine the individual client-trained (SOLO) performance without FL. We compare FedProIn with (1) FL baselines: FedAvg \cite{fedavg}, FedProx \cite{fedprox}, MOON \cite{moon}, (2) prototype-based methods: FedProto \cite{fedproto}, FedProc \cite{fedproc}, FedTGP \cite{fedtgp}, and FedPLVM \cite{fedplvm} and (3) methods that alleviate update problem: BalanceFL \cite{balancefl}. 
Since FedProIn maintains a global model, we generally exclude personalized methods that do not aggregate a model at the server. However, we make exceptions for FedProto (standard prototype based method) and FedTGP (introduces trainable global prototype). SOLO is trained for 300 local epochs, and for personalized methods (FedProto, FedTGP) that lack a global model, local predictions are aggregated for evaluation. Table~\ref{tab:results} reports the results of comparative analysis on both datasets. In both IID and non-IID settings, FedProIn outperforms all baselines on both Matek-19 dataset and HAM10000 dataset for all metrics.
Especially under non-IID conditions, FedProIn shows robustness against data heterogeneity. Especially, the existing prototype methods suffer severe performance drops across all metrics from IID to non-IID configurations. For instance, they show a decrease of 6-7\% in accuracy on HAM10000 and 1-1.6\% on Matek-19, while FedProIn restricts these relative losses to merely 2.4\% and 0.4\%, respectively. This consistent generalization is directly driven by our dual regularization (FDL and PCL), which mitigates the client drift. Conversely, personalized methods like FedProto suffer significant performance drops because their lack of a globally shared representation restricts generalization across highly skewed local data distributions. 
\begin{table}[t]
\caption{Performance comparison with the state-of-the-art on HAM10000 and Matek-19 datasets under IID and non-IID (Dirichlet $\alpha=0.5$) settings.}
\label{tab:results}
\footnotesize
\centering
\begin{tabular}{l|ccc|ccc}
\hline
\multirow{2}{*}{Method} & \multicolumn{3}{c|}{IID} & \multicolumn{3}{c}{Dirichlet (non-IID)} \\
& Acc (\%) & W-F1 (\%) & MCC & Acc (\%) & W-F1 (\%) & MCC \\
\hline
\multicolumn{7}{c}{\textbf{HAM10000}} \\
\hline
SOLO 
& 75.1 $\pm$ .3 & 73.3 $\pm$ .1 & .491 $\pm$ .002
& 61.0 $\pm$ .1 & 63.5 $\pm$ .1 & .319 $\pm$ .002 \\
\hline
FedAvg 
& 81.7 $\pm$ .2 & 80.8 $\pm$ .3 & .633 $\pm$ .005
& 78.1 $\pm$ 1.8 & 73.4 $\pm$ 2.6 & .530 $\pm$ .049 \\

FedProx 
& 81.8 $\pm$ .2 & 80.9 $\pm$ .3 & .636 $\pm$ .004
& 80.2 $\pm$ .8 & 77.9 $\pm$ .9 & .593 $\pm$ .019 \\

MOON 
& 81.5 $\pm$ 0 & 80.7 $\pm$ .1 & .629 $\pm$ .001
& 77.8 $\pm$ .8 & 73.8 $\pm$ 1.1 & .522 $\pm$ .021 \\
\hline
BalanceFL 
& 83.0 $\pm$ .7 & 82.1 $\pm$ .7 & .658 $\pm$ .014
& 79.2 $\pm$ .4 & 75.2 $\pm$ .6 & .564 $\pm$ .012 \\
\hline
FedProto 
& 73.6 $\pm$ .2 & 70.9 $\pm$ .3 & .444 $\pm$ .005
& 59.8 $\pm$ .1 & 62.0 $\pm$ .1 & .296 $\pm$ .005 \\

FedTGP 
& 62.6 $\pm$ .1 & 65.8 $\pm$ .1 & .383 $\pm$ .001
& 57.0 $\pm$ .1 & 61.2 $\pm$ .1 & .320 $\pm$ .001 \\

FedProc 
& 82.6 $\pm$ .5 & 81.9 $\pm$ .4 & .654 $\pm$ .008
& 76.4 $\pm$ .5 & 71.4 $\pm$ .8 & .483 $\pm$ .014 \\

FedPLVM
& 83.2 $\pm$ .3 &	82.6 $\pm$ .4 & .666 $\pm$ .007
& 76.2 $\pm$ .9 &	71.3 $\pm$ 2.1	& .487 $\pm$ .027 \\
\hline
FedProIn
& \textbf{83.5 $\pm$ .5} & \textbf{82.8 $\pm$ .7} & .\textbf{670 $\pm$ .013}
& \textbf{81.1 $\pm$ .3} & \textbf{79.3 $\pm$ .3} & \textbf{.613 $\pm$ .004} \\

\hline
\multicolumn{7}{c}{\textbf{Matek-19}} \\
\hline

SOLO
& 91.2 $\pm$ .3 & 90.4 $\pm$ .3 & .874 $\pm$ .004
& 77.3 $\pm$ .8 & 77.6 $\pm$ 1.0 & .688 $\pm$ .011 \\
\hline
FedAvg 
& 95.2 $\pm$ .1 & 94.7 $\pm$ .1 & .931 $\pm$ .001
& 94.8 $\pm$ .3 & 94.1 $\pm$ .3 & .925 $\pm$ .004 \\

FedProx 
& 95.1 $\pm$ .1 & 94.7 $\pm$ .1 & .930 $\pm$ .001
& 94.6 $\pm$ .1 & 93.9 $\pm$ .1 & .922 $\pm$ .002 \\

MOON 
& 95.4 $\pm$ .2 & 94.8 $\pm$ .3 & .934 $\pm$ .003
& 95.3 $\pm$ .2 & 94.6 $\pm$ .2 & .932 $\pm$ .002 \\
\hline
BalanceFL 
& 95.7 $\pm$ .3 & 95.2 $\pm$ .3 & .938 $\pm$ .004
& 95.6 $\pm$ .1 & 94.9 $\pm$ .1 & .937 $\pm$ .001 \\
\hline
FedProto 
& 91.8 $\pm$ .5 & 90.8 $\pm$ .4 & .882 $\pm$ .007
& 84.6 $\pm$ 1.6 & 83.3 $\pm$ 1.5 & .777 $\pm$ .023 \\

FedTGP 
& 71.2 $\pm$ .1 & 76.9 $\pm$ .1 & .619 $\pm$ .001
& 59.9 $\pm$ .1 & 67.1 $\pm$ .1 & .490 $\pm$ .001 \\

FedProc 
& 94.9 $\pm$ .2 & 93.8 $\pm$ .2 & .927 $\pm$ .003
& 93.3 $\pm$ 2.2 & 92.4 $\pm$ 2.3 & .904 $\pm$ .030 \\

FedPLVM
& 95.7 $\pm$ .2 &	95.2 $\pm$ .2 & .939 $\pm$ .003
& 94.7 $\pm$ .1 &	94.4 $\pm$ .1	& .927 $\pm$ .001 \\
\hline
FedProIn
& \textbf{96.2 $\pm$ .1} & \textbf{95.8 $\pm$ .1} & \textbf{.946 $\pm$ .001}
& \textbf{95.8 $\pm$ .1} & \textbf{95.1 $\pm$ .1} & \textbf{.940 $\pm$ .001} \\

\hline
\end{tabular}
\end{table}

\begin{figure}
    \centering
    % Use vector graphic format (.pdf or .eps)
    \includegraphics[width=0.9\textwidth]{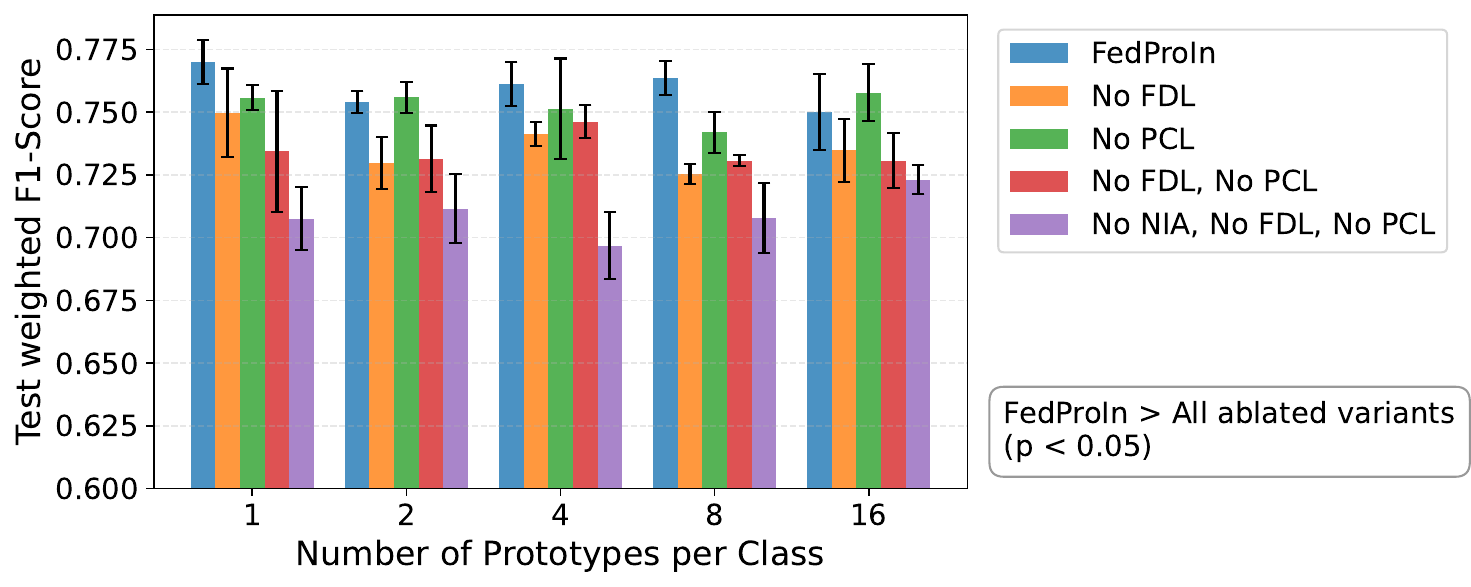} 
    \caption{\textbf{Ablation results.} Bar plots showing  weighted F1-Score with varying prototype counts ($M$) on HAM10000. Aggregated across all runs, the full FedProIn method demonstrates statistically significant improvements over all ablated variants (Mann--Whitney U test; vs. No FDL $p < 0.001$, vs. No PCL $p < 0.05$, vs. No FDL \& No PCL $p < 0.001$, vs. No NIA, FDL \& PCL $p < 0.001$).}
    \label{fig:ablation}
\end{figure}

\textit{\textbf{Ablation Study. }}
Fig.~\ref{fig:ablation} shows the ablation study results on HAM10000 under non-IID conditions with MCL. Removing FDL causes a sharper drop in W-F1 than removing PCL, confirming that feature drift is the dominant failure mode under heterogeneity. However, both components are nonetheless necessary, as the full model significantly ($p-value < 0.05$ using Mann--Whitney U test on results across prototype counts as independent samples, given our experimental design) outperforms any single-component variant. The degenerate case (no FDL, no PCL, no NIA) reduces to FedAvg and performs worst, while adding NIA alone already yields noticeable gains, validating influence-aware prototype aggregation as an effective standalone improvement.  HAM10000 favors $M=2$ prototypes per class while Matek-19 peaks at $M=1$, suggesting that the optimal prototype multiplicity is dataset-dependent.

\section{Conclusion}
In this paper, we introduce \textbf{FedProIn}, a multiple Learnable Prototype based framework to tackle client drift by decomposing it into feature drift and prototype drift. We mitigate the feature and prototype drift by introducing Feature divergence loss and prototype contrastive loss. We also introduce a novel aggregation method for learnable prototypes, which  prioritizes prototypes based on their local utility. Experimental results on the HAM10000 and Matek-19 datasets demonstrate that FedProIn outperforms existing federated learning methods. Future work will explore adapting FedProIn for personalized FL by eliminating the dependency on a global model to enhance local adaptability. We also aim to extend the framework to heterogeneous FL settings, enabling collaboration across clients with diverse model architectures.

\paragraph{\textbf{Acknowledgments. }}This work was supported by DBT/Wellcome Trust India Alliance Fellowship [IA/E/22/1/506763], Council of Scientific \& Industrial Research (CSIR) ASPIRE program [25WS(013)/2023-24/EMR-II/ASPIRE], Start-up Research Grant [SRG/2023/001406] from the Science and Engineering Research Board  and Pratiksha Trust, Bangalore, India [FG/PTCH-23-1004].
 
\paragraph{\textbf{Disclosure of Interests. }}The authors declare no conflict of interest.


\begin{thebibliography}{8}


\bibitem{guan2024fl}
Guan, H., Yap, P.-T., Bozoki, A., Liu, M.: Federated learning for medical image analysis: A survey. Pattern Recognition \textbf{151}, 110424 (2024). \doi{10.1016/j.patcog.2024.110424} 


\bibitem{yang_cvpr_2018}
Yang, H.-M., Zhang, X.-Y., Yin, F., Liu, C.-L.: Robust Classification With Convolutional Prototype Learning. In: Proceedings of the IEEE Conference on Computer Vision and Pattern Recognition (CVPR), pp. 3474--3482. IEEE, Salt Lake City, UT, USA (2018). \doi{10.1109/CVPR.2018.00366}

\bibitem{balancefl}
Shuai, X., Shen, Y., Jiang, S., Zhao, Z., Yan, Z., Xing, G.: BalanceFL: Addressing Class Imbalance in Long-Tail Federated Learning. In: 2022 21st ACM/IEEE International Conference on Information Processing in Sensor Networks (IPSN), Milano, Italy, pp. 271–284. IEEE (2022). \doi{10.1109/IPSN54338.2022.00029}

\bibitem{hipaa}
HHS.gov: HIPAA Home | HHS.gov, https://www.hhs.gov/hipaa/index.html, last accessed 2025/01/12.

\bibitem{gdpr}
GDPR Info: General Data Protection Regulation (GDPR). \url{https://gdpr-info.eu/}, last accessed 2026/01/25


\bibitem{fedavg}
McMahan, B., Moore, E., Ramage, D., Hampson, S., Arcas, B.A.: Communication-efficient learning of deep networks from decentralized data. In: Proceedings of the 20th International Conference on Artificial Intelligence and Statistics (AISTATS), pp. 1273--1282 (2017)

\bibitem{fedprox}
Li, T., Sahu, A.K., Zaheer, M., Sanjabi, M., Talwalkar, A., Smith, V.: Federated optimization in heterogeneous networks. In: Proceedings of Machine Learning and Systems (MLSys), vol. 2, pp. 429--450 (2020). \url{https://arxiv.org/abs/1812.06127}

\bibitem{moon}
Li, Q., He, B., Song, D.: Model-contrastive federated learning. In: Proceedings of the IEEE/CVF Conference on Computer Vision and Pattern Recognition (CVPR), pp. 10709--10719 (2021). \doi{10.1109/CVPR46437.2021.01057}

\bibitem{fedproto}
Tan, Y., Long, G., Liu, L., Zhou, T., Lu, Q., Jiang, J., Zhang, C.: FedProto: Federated Prototype Learning across Heterogeneous Clients. In: Proceedings of the AAAI Conference on Artificial Intelligence, vol. 36, no. 8, pp. 8432--8440. AAAI Press (2022). \doi{10.1609/aaai.v36i8.20819}

\bibitem{fedproc}
Mu, X., Shen, Y., Cheng, K., Geng, J., Jia, J., Li, T.: Fedproc: Prototypical contrastive federated learning on non-iid data. Future Generation Computer Systems \textbf{143}, 93--104 (2023)


\bibitem{fedtgp}
Zhang, J., Liu, Y., Hua, Y., Cao, J.: FedTGP: Trainable global prototypes
with adaptive-margin-enhanced contrastive learning for data and model
heterogeneity in federated learning. In: Proceedings of the AAAI Conference
on Artificial Intelligence, vol. 38, no. 15, pp. 16768--16776. AAAI Press (2024)


\bibitem{fpl}
Huang, W., Li, J., Chen, Y., Ding, Z., Zhou, Z.-H.:
Rethinking federated learning with domain shift: A prototype view.
In: Proceedings of the IEEE/CVF Conference on Computer Vision and Pattern Recognition (CVPR 2023),
pp. 16312--16321. IEEE (2023)

\bibitem{fedplvm}
Wang, L., Zhang, Y., Li, X., Jin, D., Yang, Q.:
Taming cross-domain representation variance in federated prototype learning with heterogeneous data domains.
In: Advances in Neural Information Processing Systems,
vol.~37, pp. 88348--88372 (2024)


\bibitem{ham10000}
Tschandl, P., Rosendahl, C., Kittler, H.: The HAM10000 dataset, a large collection of multi-source dermatoscopic images of common pigmented skin lesions. Scientific Data \textbf{5}(1), 1--9 (2018)

\bibitem{Matek-19}
Matek, C., Schwarz, S., Spiekermann, K., et al.: Human-level recognition of blast cells in acute myeloid leukaemia with convolutional neural networks. Nat. Mach. Intell. \textbf{1}, 538--544 (2019). \doi{10.1038/s42256-019-0101-9}

\bibitem{scaffold}
Karimireddy, S.P., Kale, S., Mohri, M., Reddi, S., Stich, S.U., Suresh, A.T.: 
Scaffold: Stochastic controlled averaging for federated learning. 
In: III, H.D., Singh, A. (eds.) Proceedings of the 37th International Conference on Machine Learning (ICML 2020), 
Proceedings of Machine Learning Research, vol. 119, pp. 5132--5143. PMLR (2020)

\bibitem{feddyn}
Acar, D.A.E., Zhao, Y., Navarro, R.M., Mattina, M., Whatmough, P.N., Saligrama, V.: 
Federated learning based on dynamic regularization. 
In: International Conference on Learning Representations (ICLR) (2021)

\bibitem{fedpcl}
Tan, Y., Long, G., Ma, J., Liu, L., Zhou, T., Jiang, J.: 
Federated learning from pre-trained models: A contrastive learning approach. 
In: Koyejo, S., Mohamed, S., Agarwal, A., Belgrave, D., Cho, K., Oh, A. (eds.) 
Advances in Neural Information Processing Systems 35 (NeurIPS 2022), 
pp. 19332--19344 (2022)

\bibitem{mcc}
Matthews, B.W.: Comparison of the predicted and observed secondary structure of T4 phage lysozyme. Biochim. Biophys. Acta \textbf{405}(2), 442--451 (1975)

\bibitem{imagenet}
Deng, J., Dong, W., Socher, R., Li, L.-J., Li, K., Fei-Fei, L.: 
ImageNet: A large-scale hierarchical image database. 
In: 2009 IEEE Conference on Computer Vision and Pattern Recognition (CVPR), 
pp. 248--255. IEEE (2009)

\bibitem{adam}
Kingma, D.P., Ba, J.: Adam: A method for stochastic optimization. arXiv preprint arXiv:1412.6980 (2014)

\bibitem{paszke2019pytorch}
Paszke, A and Gross, S and Massa, F and Lerer, A., et al.: Pytorch: An imperative style, high-performance deep learning library. Advances in neural information processing systems \textbf{32}, 2019


\end{thebibliography}
%

\end{document}